\documentstyle[aps,epsf,graphics]{revtex}
\def\br{\begin{eqnarray}}
\def\er{\end{eqnarray}}
\def\be{\begin{equation}}
\def\ee{\end{equation}}

\def\L{\Lambda}

\def\({\left(}
\def\){\right)}

\begin{document}

%\twocolumn[\hsize\textwidth\columnwidth\hsize\csname %%% TWO COLUMN
%@twocolumnfalse\endcsname                            %%% TWO COLUMN
%

%\draft
%
\title{Phenomenological tests for the freezing of the QCD running coupling constant}
\author{A.~C.~Aguilar, A.~Mihara and A.~A.~Natale\\}
\address{Instituto de F\'{\i}sica Te\'orica,
UNESP, Rua Pamplona 145, 01405-900, S\~ao Paulo, SP, Brazil}
%
%%%
\date{\today}
\maketitle
%%%%

\begin{abstract}
We discuss phenomenological tests for the frozen infrared behavior of the running coupling
constant and gluon propagators found in some solutions of Schwinger-Dyson equations of the
gluonic sector of QCD. We verify that several observables can be used in order to select
the different expressions of $\alpha_s$ found in the literature. We test the effect of the
nonperturbative coupling in the $\tau$-lepton decay rate into nonstrange hadrons, in the $\rho$
vector meson helicity density matrix that are produced in the $\chi_{c2}\rightarrow \rho\rho$ decay,
in the photon to pion transition form factor, and compute the cross sections for
elastic proton-proton scattering and exclusive $\rho$ production in deep inelastic
scattering. These quantities depend on the infrared behavior of the coupling
constant at different levels, we discuss the reasons for this dependence and argue that the existent
and future data can be used to test the approximations performed to solve the Schwinger-Dyson
equations and they already seems to select one specific infrared behavior of the coupling.
\end{abstract}


\vskip 0.5cm                            %two column

\section{Introduction}

The interface between perturbative and non\-per\-tur\-ba\-tive QCD has been
studied for many years. To know how far we can go with perturbative
calculations and how we can match these ones with nonperturbative quantities,
obtained through sophisticated theoretical or phenomenological approaches,
will probably still require several years of confront between these methods.
However there are phenomenological indications that this connection may
not be abrupt and the transition from short distance (where quark and gluons
are the effective degrees of freedom) to long distance (hadron) physics is
indeed a smooth one \cite{dokshitzer}. Actually it has been suggested that the
strong coupling constant freezes at a finite and moderate value \cite{stevenson},
and this behavior could be the reason for the claimed soft transition.
The freezing of the QCD running coupling at low energy scales could allow
to capture at an inclusive level the nonperturbative QCD effects in a
reliable way \cite{dokshitzer,brodsky0}.

The aspects described above reflect a perspective to attack the
short and long-distance QCD interface from the perturbative side. On the other
direction we have nonperturbative methods to investigate the infrared QCD
behavior. One of these methods, based on the study of gauge-invariant
Schwinger-Dyson equations, concluded that an infrared finite coupling
constant could be obtained from first principles \cite{cornwall}. A
list of nonperturbative attempts to determine the
infrared behavior of the gluon coupling constant and propagator can
be found in Ref.\cite{we}, where it is also pointed out that the
nonperturbative results for these quantities may differ among themselves
due to the intricacies of the nonperturbative methods and to not always
controllable approximations.

Recently new theoretical results about the infrared behavior of the gluon
propagator and the running coupling constant appeared in the literature. We
have a renormalization group analysis implying in an infrared finite
coupling \cite{gies}, and also new numerical lattice studies simulating the
gluon propagator, all of them consistent with an infrared finite behavior \cite{lat}.
But the most interesting for us are the new solutions for the gluon
and ghost Schwinger-Dyson equations that have been obtained with better
approximations \cite{bloch,alkofer,smekal,zwanziger}. The absence of certain angular approximations
in the integrals of the Schwinger-Dyson equations (SDE) are some of the
improvements in the Refs.\cite{alkofer,smekal}, which led to a value
for the infrared coupling constant at the origin ($\alpha_s(0)$) roughly
a factor $O(3)$ below the previous ones \cite{alkofer0}.

Assuming that the coupling constant and the gluon
propagator are infrared finite we can divide the Schwinger-Dyson solutions
for the gluon propagator in two classes. One where the gluon propagator is
identical to zero at the momentum origin \cite{bloch,alkofer,smekal,zwanziger}
and another where the propagator is roughly of order $1/{m_g}^2$\cite{cornwall},
where $m_g$ is the dynamical gluon mass. In Ref.\cite{we}
the phenomenological calculation of the asymptotic pion form factor was
used to show that the experimental data seems to prefer the Cornwall's
solution for the coupling and gluon propagator\cite{cornwall}.

In this work we will extend our approach \cite{we}
to others observable quantities, testing
the recent nonperturbative results obtained for $\alpha_s$ and for the
gluon propagators. These tests span from a purely perturbative calculation
to cross sections containing some model dependence.
We verify that these quantities are affected by the
infrared behavior of the coupling in different amounts. However we can
certainly see that the existent data can be used to test and distinguish
different nonperturbative calculations of the infrared QCD behavior.

There are several important points to stress. The SDE solutions represent
attempts of a very ambitious program aiming to obtain the behavior of the
coupling constant at all scales. Of course, these equations contain approximations and there
are even questionings about the definition of the coupling in the infrared, but once these
facts are assumed and knowing that in Nature there must be only one coupling constant
we should be able to compare these results to the data. Some of the tests we propose here
are purely perturbative, but they still can say something about the approximations used to
obtain $\alpha_s$. Another point is that the
whole idea of a running charge is that it is to be used in a skeleton expansion
of graphs representing a process, an expansion which automatically subsumes
infinitely many graphs, in a scheme emphasized by Brodsky and collaborators \cite{brodsky1}.
But if, contrary to many phenomenological evidences
\cite{dokshitzer,brodsky0,brodsky1}, the infrared value of the coupling constant
 were large enough (how large depends on the process), it is hard to see how one can make sense
out of inserting such a large value into perturbative expressions.  This point is at the heart of the discussions
that we shall present in this paper.  The SDE solutions are determined forcing its high
energy behavior to be identical to the perturbative one, after this is done the infrared behavior
is totally dictated by the equations. This procedure is a natural one because
at large momenta the coupling constant is perturbative and the QCD determination of
the leading S matrix terms is meaningful, although the perturbative quantities depend
on an infrared parametrization hidden in the QCD scale ( $\Lambda_{QCD}$).

The solutions that we shall discuss have different values for $\alpha_s(0)$.
It could be argued that the value of $\alpha_s(0)$
for one of these SDE solutions is too small to explain chiral symmetry breaking in the usual model with
essentially one-gluon exchange \cite{silva}. It has been known for many years that
this sort of model of chiral symmetry breaking doesn't work, precisely because
it does require such large values for $\alpha_s(0)$. In QCD, unlike QED, chiral
symmetry breaking is easily explained --indeed, required by-- confinement, as
numerous authors in the eighties have explained in detail. A value of $\alpha_s (0)$
of order one or larger would be grotesquely out of line with many other
phenomena, such as non-relativistic potentials for charmonium \cite{simonov}.
Finally, this is also what comes out when performing an analytic perturbation
theory \cite{shirkov}. Moreover, the fact is that the nonperturbative solutions
do not have only different  $\alpha_s(0)$ values but they also run differently near the
infrared. The different infrared parametrization and running will leave a signal
in the comparison with the experimental data.

The distribution of the paper is the following: In Section II we present
a comparison between the most recent calculations of the infrared behavior
of the QCD running coupling constant. Section III compares the nonperturbative
value of $\alpha_s$ with its value measured in $\tau$-lepton decay into
nonstrange hadrons. In Section IV we discuss the
effects of a frozen coupling constant in the diagonal elements of the
$\rho$ vector meson helicity density matrix that are produced in
the $\chi_{c2}\rightarrow \rho\rho$ decay. Section V is devoted to
a discussion of the $\gamma \rightarrow \pi_{0}$ transition form factor.
In Section VI we revisit the calculation of Ref.\cite{halzen}, where the
Pomeron model of Landshoff and Nachtmann for elastic proton-\-proton scattering
was used to restrict the value of the dynamical gluon mass, in the
light of the new SDE solutions. The same model is used in Section VII to
compute the exclusive $\rho$ production in deep inelastic scattering.
Section VIII contains our conclusions.

\section{Infrared Behavior of the Running Coupling Constant}

In this Section we present the results for the running
coupling constant obtained recently through the solution of SDE for
the gluon and ghost sectors of QCD. As remarked in Ref.\cite{we} and
in our introduction, it is important to stress that the different
solutions appear due to the different approximations made to solve
the SDE, which, unfortunately, are necessary due to their complicated
structure.

The most recent calculations of the $\alpha_s$
infrared behavior do not make use of an approximate
angular integration in the SDE, resulting in a value of $\alpha_s(0)$ a
factor of $O(3)$ below the former results.
This is the case of the $\alpha_s(q^2)$ determined by Fischer and
Alkofer which is given by \cite{alkofer}

\begin{equation}\label{runalk}
  \alpha_{sA} (x)= \frac{\alpha_A(0)}{\ln (e + a_1x^{a2} +b_1x^{b2}) } ,
\end{equation}

\noindent
where

\hskip 0.5in$  \alpha_A (0) = 2.972$,

\hskip 0.5in$  a_1= 5.292\;\mbox{GeV}^{-2a_2}$,

\hskip 0.5in$  a_2=2.324$,

\hskip 0.5in$  b1= 0.034\;\mbox{GeV}^{-2b_2}$,

\hskip 0.5in$  b2=3.169$.

We also have an ansatz for the running coupling proposed by
Bloch. Its expression for
$\alpha_s(q^2)$ has the following fit \cite{bloch}:

\begin{eqnarray}\label{bloch}
\alpha_{sB}(q^2)&=& \alpha(l\Lambda_{QCD}^2) = \\ \nonumber
&&\frac{1}{c_{0} + l^2}\left[ c_{0}\alpha_0 +
\frac{4\pi}{\beta_0}\left( \frac{1}{\log(l)} -
\frac{1}{l-1}\right) l^2 \right]
\end{eqnarray}

\noindent
where $l=q^2/\Lambda^2_{QCD}$, $c_{0} = 15$, $\alpha_0 = 2.6$, and
$\beta_0 = 11 - \frac{2}{3}n_f$, where $n_f$ is the number of flavors.

\begin{figure}[ht]
{\par
\centering 
\resizebox*{12.14cm}{10.0cm}
{\includegraphics{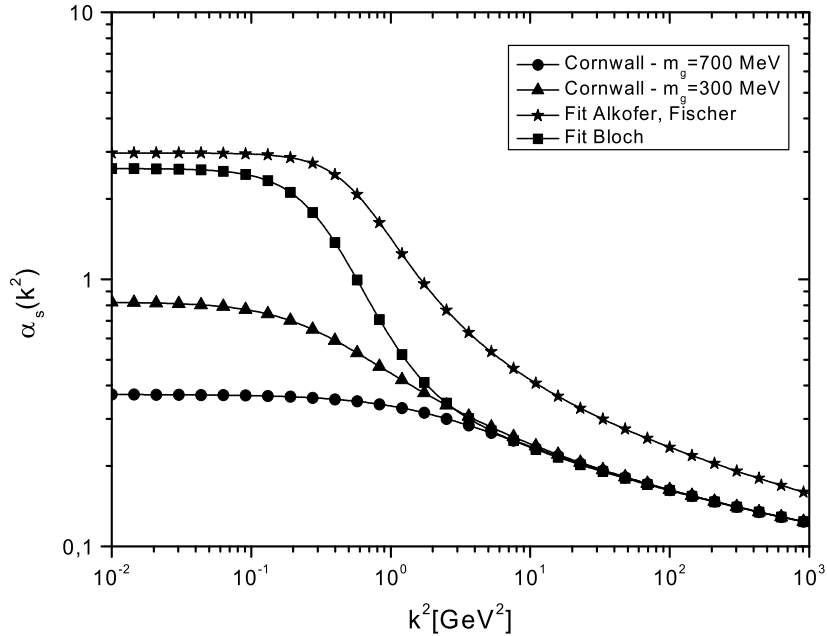}} 
\par} 
\caption[dummy0]{Comparison between the running couplings
obtained from different solutions of Schwinger-Dyson equations.} 
\label{running}
\end{figure}

Equation (1) has been obtained solving the coupled SDE for the
gluon and ghost propagators in the Landau gauge. Eq. (2) is an
ansatz motivated by earlier results done with the help of the
angular approximation. Both are consistent with a gluon propagator
that vanishes in the infrared. The difference in the couplings,
which is clearly displayed in Fig.(\ref{running}), is that Bloch
fix the scale of the coupling constant at $\Lambda_{QCD}=300$MeV
while Alkofer and Fischer \cite{alkofer} fix it comparing to the
experimental input at $\alpha_s (M_Z)$, unfortunately their
calculation when compared to a perturbative expression for
coupling constant in the MOM scheme lead to a value $\Lambda_{QCD}=715$MeV
which seems to be too large. The difference between the values of $\alpha_s(0)$
is negligible, while the difference at intermediate momenta is
going to be transferred to the phenomenological quantities that we
will compute.

%%%%%%%%%%%%%%%%%%%%%%%%%%%%%%%%%%%%%%%%%%%%%%%%%%%%%%%%%%%%%%%%%%%%
%\begin{figure}[ht]
%\setlength{\epsfxsize}{1.0\hsize}
%\centerline{\epsfbox{fig1.eps}}%\vskip 1.0 cm

%%%%%%%%%%%%%%%%%%%%%%%%%%%%%%%%%%%%%%%%%%%%%%%%%%%%%%%%%%%%%%%

The other expression for the running coupling constant that we will use in the
next sections is the one determined by Cornwall many years ago \cite{cornwall}
\be \alpha_{sC} (q^2)= \frac{4\pi}{\beta_0 \ln\left[
(q^2 + 4M_g^2(q^2) )/\L^2 \right]}, \label{acor} \ee
where $M_g(q^2)$ is a dynamical gluon mass given by,
\be M^2_g(q^2) =m_g^2 \left[\frac{ \ln
\(\frac{q^2+4{m_g}^2}{\Lambda ^2}\) } {
\ln\(\frac{4{m_g}^2}{\Lambda ^2}\) }\right]^{- 12/11}
\label{mdyna} \ee
$\L$($\equiv\L_{QCD}$) is the QCD scale parameter. This solution is the only
one that has been obtained in a gauge invariant procedure. The gluon mass scale has
to be found phenomenologically. A typical value
is \cite{cornwall,halzen}
\begin {equation}
m_g = 500 \pm 200 \quad {\textnormal MeV}
\label{mg}
\end{equation}
for $\L = 300$ MeV. The Bloch and Cornwall's expression at large momenta
map perfectly into the perturbative running coupling, as can be seem in
Fig.(\ref{running}).  We will refer to these different results as model A, B
and C respectively.

\section{$\tau$--Lepton Decay Rate into Nonstrange Hadrons}

The tests of the running coupling constant behavior that we shall
discuss will depend on the exchange of gluons at different levels, i.e.
the softer is the exchanged gluon the more we can test the
nonpertubative behavior of $\alpha_s$. The physics of the $\tau$-lepton
decay rate into nonstrange hadrons is one where QCD can be confronted
with experiment to a very high precision. The measurement of $\alpha_s$
in these decays and its comparison with the nonperturbative expressions
will correspond to what can be called our most perturbative test.

The normalized $\tau$-lepton decay rate into nonstrange hadrons ($h_{S=0}$)
is given by \cite{korner}
\begin{eqnarray}
\label{eq:rtau}
\left. R(\tau )\right| _{S=0} &=&
\frac{\Gamma (\tau \rightarrow h_{S=0}\nu )}
{\Gamma (\tau \rightarrow l\overline{\nu }\nu )} \nonumber\\
&=&N_{c}\left| V_{ud}\right| ^{2}S_{ew}(1+\delta _{ew}+\delta _{P}+\delta _{NP}),
\end{eqnarray}
where $ N_{c} $ is the number of colors, the flavor mixing matrix element
is $ \left| V_{ud}\right| ^{2}=0.9511\pm 0.0014 $ and $ S_{ew}=1.0194 $
and $ \delta _{ew}=0.001 $ are electroweak corrections. The nonperturbative
corrections are small and consistent with zero, $ \delta _{NP}=-0.003\pm 0.003 $.
In particular, these corrections are not directly related to the nonperturbative
infrared behavior of $\alpha_s$.
The term $ \delta _{P} $ represents the perturbative QCD effects and its
value can be calculated if we have $ \left. R(\tau )\right| _{S=0} $. With
the experimental result obtained by the ALEPH Collaboration\cite{aleph},
$ \left. R(\tau )\right| ^{expt}_{s=0} $=3.492$ \pm 0.016 $,
and the other values above one can estimate the perturbative corrections:
\begin{equation}
\label{eq:dpexp}
\delta ^{expt}_{P}=0.203\pm 0.007.
\end{equation}
On the other hand $ \delta _{P} $ can be calculated in the framework of perturbative
QCD and, in the $\overline{\mbox{MS}}$ scheme, is given by the expansion
\begin{equation}
\label{eq:dpth}
\delta ^{th}_{P}=\left( \frac{\alpha _{s}}{\pi }\right)
+5.2023\left( \frac{\alpha _{s}}{\pi }\right) ^{2}+26.366\left( \frac{\alpha _{s}}{\pi }\right) ^{3}+\cdots \, ,
\end{equation}
where $ \alpha _{s} $ is the running coupling constant taken at the scale
of the $ \tau  $-lepton mass $ M_{\tau }=1.777 $GeV. Substituting Eq.(\ref{eq:dpexp})
in the left hand side of Eq.(\ref{eq:dpth}) and solving the resulting polynomial equation
in $ \alpha _{s} $ one obtain
\begin{equation}
\label{eq:alpha_exp}
\alpha ^{expt}_{s}=0.3404\pm 0.0073.
\end{equation}

Now we can compare the above experimental result with the different values of
the nonperturbative running coupling at the $ \tau  $-lepton mass scale.  Some
comments are in order before such comparison is performed. The nonperturbative couplings
have been obtained in the case of zero flavors. When we compare the perturbative to
the nonperturbative results it should be understood that  the introduction of fermions will affect the
three models in the same direction (increasing $\alpha_s(0)$) and this effect would be small for a
small number of flavors \cite{cornwall,we}. Another point is that Eq.(\ref{eq:dpth})
has a renormalization scheme dependence, while the nonperturbative solutions have
been obtained in one unrelated approach. However the nonperturbative couplings of the
three models are obtained forcing the matching with the perturbative running coupling at very large momenta and at
leading order there is no problem with scheme dependence, in such a way that the comparison can be
made but not with the precision of the experimental result.
Using Eqs.~(\ref{runalk}) to (\ref{acor}) (with $m_g =500$ MeV and $\Lambda =300$ MeV)
instead of the perturbative running coupling we obtain
\begin{equation}
\alpha_{sA}=0.676\; ,\quad \alpha_{sB}=0.354\; ,\quad \alpha_{sC}=0.348.
\end{equation}
As one could expect only the last two results are compatible with Eq.(\ref{eq:alpha_exp}), and
the model A fail at a level much above the one where we could claim that any scheme dependence
is important. This shows the importance of the scale used to fix the coupling constant. Note that
Alkofer and Fischer \cite{alkofer} (model A) constrain the coupling comparing its value to the
experimental one of $\alpha_s (M_Z)$, which seems to be the most rigorous way of
doing it, unfortunately their calculation lead to a value $\Lambda_{QCD}=715$MeV,
which is too high when compared to most of the values found in the literature.
As  the $ \tau  $-lepton mass scale is one where we believe that perturbative QCD
can already be applied we have to conclude that the SDE solutions of model A does
not describe the experimental data at such energies. Model B does it quite well but
it was fitted to a smaller  $\Lambda_{QCD}$ value.  Note that the model A does not
fit the data not because it has a large value for $\alpha_s(0)$, which is not so different
from the model B, but because the running coupling of model A decreases much more
slowly.

As we claimed previously, this test is the most perturbative one
in the sense that it does not involve the introduction of hadronic wave functions directly in
the calculation. In the next sections we always will need to introduce some extra hadronic
information besides the coupling and gluon propagator. It is important to stress that
this  result does not settle the question about which one is the correct solution of gluonic
SDE, it only give hints to which approximations may be or not suitable to be performed
when solving these equations.

\section {Effects of a frozen coupling constant in the amplitudes
of the $\chi_{c2}\rightarrow \rho\rho$ decay}

It is known, for a long time, that a dynamical gluon mass can
affect quarkonium decays \cite{cornwall,PP,mihara}. In the study
of Refs. \cite{PP,mihara} it can be noted that the dynamical mass
effect is relevant in the study of their hadronic decays mostly
due to phase space factors. Therefore not all quarkonium decays
are good candidates to probe the infrared behavior of the running
coupling constant, because the effects of phase space overwhelms
the changes in the running coupling.

It is also known that the exclusive quarkonium production
in $p\bar{p}$ or $pp$ interactions and subsequent decay is strongly dependent on the
internal structure of the hadrons involved, which are described by their hadronic
wave functions. In the inclusive case the important dependence is on the distribution
function. Perturbative QCD fixes the asymptotic form of these functions as $q^2 \rightarrow \infty$
and their general evolution, but in any realistic calculation they have to be
taken as phenomenological quantities, to be experimentally determined via a set
of physical information and then used in other processes.

Even when we are dealing with heavy quarks, and try to propose tests for the
interface between perturbative and non-per\-tur\-bative QCD, we verify that
we cannot test the infrared behavior of quarks and gluons vertex and propagators
independently of hadronic wave or distribution functions. In this Section we discuss the effects of a
frozen coupling constant in the measurement of the $\rho$ vector meson helicity
density matrix that are produced in the $\chi_{c2}\rightarrow \rho\rho$ decay.
This is a case where the infrared behavior of the coupling is entangled with
the behavior of the wave or distribution functions. We advance that in Section V we will discuss
a more ordinary case where the wave function is well known, and where it is possible
to study the QCD infrared behavior with less uncertainty \cite{halzen}.

Some time ago Anselmino and Murgia \cite{murgia} considered the $\chi_{c2}\rightarrow \rho\rho$ decay
process of polarized charmonium states created in  $p\bar{p}$ or $pp$ interactions and
have shown how the observation of the polarization of the vector meson, via a measurement
of its diagonal helicity density matrix elements, neatly depends on the $\rho$ wave function
and helps in discriminating between different kinds of these quantities.

The study of Ref.\cite{murgia} provide an interesting arena to introduce the effects of
a frozen running coupling constant. Our result will show that depending on the form of
the wave function we could observe a larger or smaller effect of the freezing in the infrared,
showing how the nonpertur\-bative behaviors of wave (or distribution) functions with the
ones of the effective coupling can become entangled.

The processes that we consider are the exclusive

\begin{equation}\label{exclusive}
p\overline{p}\rightarrow \chi_{c2} \rightarrow \rho\rho
\end{equation}

\noindent
or the inclusive

\begin{equation}\label{inclusive}
pp \rightarrow \chi_{c2} + X \rightarrow \rho\rho + X
\end{equation}

\noindent
production of a pair of $\rho$ vectors with the subsequent decay

\begin{equation}\label{decay}
\rho \rightarrow \pi\pi ,
\end{equation}

\noindent
and the quantity experimentally observed is the angular distribution of either one of the
pions in the helicity rest frame of the decaying $\rho$.

The pion angular distribution depends of the spin state of the $\rho$ via the elements of
its helicity density matrix $\rho_{\lambda,\lambda'} (\rho)$:

\begin{eqnarray}\label{wpi}
W (\Theta,\Phi) &=&  \frac{3}{4\pi}
[\rho_{0,0} \cos^2{\Theta} +(\rho_{1,1}- \rho_{1,-1})\sin^2{\Theta}\cos^2{\Phi}
+ (\rho_{1,1}+ \rho_{1,-1})\sin^2{\Theta}\sin^2{\Phi}
\nonumber \\
&&-\sqrt{2} (Re\rho_{1,0}) \sin 2{\Theta}\cos{\Phi} ],
\end{eqnarray}

\noindent
where $\Theta$ and $\Phi$ are, respectively, the polar and azimuthal angles of the
pion as it emerges from the decay of the $\rho$, in the $\rho$ helicity rest frame.
Eq. (\ref{wpi}) can be integrated over $\Theta$ or $\Phi$ generating polar and
azimuthal distributions, which can be measured and give information on
$\rho_{\lambda,\lambda'} (\rho)$. The details of this procedure (i.e. the vector
meson helicity density matrix calculation for massless quarks) can be found
in Ref.\cite{murgia} and references therein.

The helicity density matrix of the $\rho$ meson are \cite{murgia}

\begin{equation}\label{densmatr00}
\rho_{\lambda,\lambda'}= 0, \,\,\,\,\, \lambda \neq \lambda', \,\,\,\,\,\,\,\,\,\,\,\,\,\,\,\,\,\,\,
\rho_{0,0} = 1 -2 \rho_{1,1}
\end{equation}

\begin{equation}\label{densmatrix}
\rho_{1,1}=\rho_{-1,-1} = \frac{1}{2} \frac{1}{1+
 3\frac{|\mbox{\~A}_{0,0}|^2}{|\mbox{\~A}_{1,-1}|^2}F(\theta)},
\end{equation}

\noindent
where the reduced amplitudes ${\tilde{A}}_{\lambda,\lambda'}$ are given in Eqs. (2.10),
(2.11) and (2.14), (2.15) of Ref.\cite{anselmino} and do not depend on the
$\rho$ production angles $\theta$ and $\phi$, but do depend on the wave functions.
Eq. (\ref{densmatrix}) has the same form both for exclusive and inclusive $\rho$
production, but the dependence on the production angle $\theta$ is different in the
two cases

\begin{equation}\label{angleex}
  F^{ex}(\theta)= \frac{\cos^2(\theta)}{1+ \cos^2(\theta)}   ,
\end{equation}

\begin{equation}\label{anglein}
F^{in}(\theta)=\frac{\sin^4(\theta)}{1+6\cos^2(\theta) +
\cos^4(\theta)}.
\end{equation}

The ratio of reduced amplitudes $|\mbox{\~A}_{0,0}|^2/|\mbox{\~A}_{1,-1}|^2$ appearing
in Eq.(\ref{densmatrix}) are computed as a function of the $\rho$ longitudinally $(L,\lambda=0)$
and transversely  $(T,\lambda= \pm 1)$ polarized vector mesons wave functions, which are
indicated respectively by $\varphi_{L}$ and $\varphi_{T}$, and correspondent decay constants given by
$f_L$ and $f_T$. This ratio is equal to \cite{murgia}

\begin{equation}\label{ratea}
\frac{|\mbox{\~A}_{0,0}|}{|\mbox{\~A}_{1,-1}|}=
\frac{1}{\sqrt6}\left(\frac{f_L}{f_T}\right)^2
\frac{|I_{0,0}|}{|I_{1,-1}|}
\end{equation}

\noindent
where

\begin{eqnarray}\label{i1-1}
I_{1,-1} = - \frac{1}{32}\int ^{1}_{0}\,
dxdy\varphi_{T}(x,\mbox{\~{Q}}_x^2)\varphi_{T}(y,\mbox{\~{Q}}_y^2)
%\nonumber \\
%\times
\frac{\alpha_s(xyM^2_{\chi})\alpha[(1-x)(1-y)M^2_{\chi}]}{xy(1-x)(1-y)(2xy-x-y)}
\end{eqnarray}

\begin{eqnarray}\label{i00}
I_{0,0} = &-& \frac{1}{32}\int ^{1}_{0}\,
dxdy\varphi_{L}(x,\mbox{\~{Q}}_x^2)\varphi_{L}(y,\mbox{\~{Q}}_y^2)
%\nonumber \\
%&&\times
\frac{\alpha_s(xyM^2_{\chi})\alpha[(1-x)(1-y)M^2_{\chi}]}
{xy(1-x)(1-y)(2xy-x-y)}
%\nonumber \\
%&&\times 
\left[ 1+ \frac{(x-y)^2}{2xy -x -y }\right].
\end{eqnarray}

\noindent
In the above equations $M_{\chi}= 3.6 \,\mbox{GeV}$ is the $\chi_{c2}$ mass and
$\mbox{\~{Q}}_x= \min(x,1-x)Q$.

To calculate the helicity density matrix $\rho_{1,1}$ we choose two different sets
of $\varphi_{L}$ and $\varphi_{T}$: a set of symmetric distribution amplitudes

\begin{equation}\label{symda}
\varphi_{L}= \varphi_{T}=6x(1-x)
\end{equation}

\noindent
and the QCD sum rules amplitudes \cite{cher}

\begin{eqnarray}\label{sumrulesl}
\varphi_{L}(x,&\mbox{\~{Q}}_{x}^2&)= 6x(1-x) 
%\nonumber \\
%&\times& 
\left\{1+\frac{1}{5}C_{2}^{3/2}(2x-1)
\left[\frac{\alpha_s(\mbox{\~Q}^2_x)}{\alpha_s(\mu_{L}^2)}\right]^{2/3}
\right\}
\end{eqnarray}

\begin{eqnarray}\label{sumrulest}
\varphi_{T}(x,\mbox{\~{Q}}_{x}^2)= 6x(1-x)
\left\{\left[\frac{\alpha_s(\mbox{\~Q}^2_x)}{\alpha_s(\mu_{T}^2)}\right]^{4/25}
 \right. 
%\\ \nonumber
\left. -\frac{1}{6}
C_{2}^{3/2}(2x-1)\left[\frac{\alpha_s(\mbox{\~Q}^2_x)}{\alpha_s(\mu_{T}^2)}\right]^{52/75}
\right\}
\end{eqnarray}

where $\mu_{L}^2=0.5 \, (\mbox{GeV}/c)^2$, $\mu_{T}^2=0.25\,
(\mbox{GeV}/c)^2$, and $C_{z}$ denotes Gegembauer polynomials. In both cases we
assume $f_L = f_T$.

%%%%%%%%%%%%%%%%%%%%%%%%%%%%%%%%%%%%%%%%%%%%%%%%%%%%%%%%%%%%%%%%%%%%
%\begin{figure}[ht]
%\setlength{\epsfxsize}{1.0\hsize}
%\centerline{\epsfbox{fig2.eps}}%\vskip 1.0 cm
\begin{figure}[ht]
{\par
\centering 
\resizebox*{10.9cm}{9.0cm}
{\includegraphics{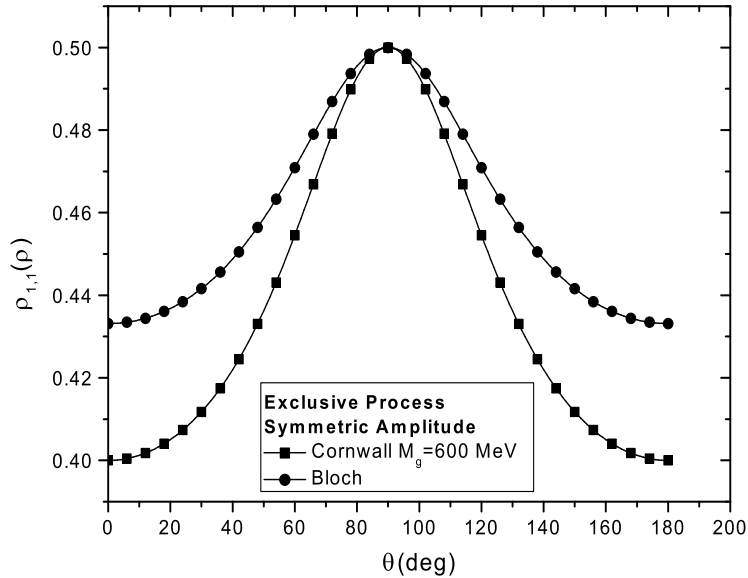}} 
\par}
\caption[dummy0]{Values of the helicity matrix element $\rho_{1,1}$ as a function of
the $\rho$-meson production angle $\theta$. The exclusive process is computed with the symmetric
distribution amplitudes for the Bloch and Cornwall's coupling constants.} \label{exsym}
\end{figure}
%%%%%%%%%%%%%%%%%%%%%%%%%%%%%%%%%%%%%%%%%%%%%%%%%%%%%%%%%%%%%%%

%%%%%%%%%%%%%%%%%%%%%%%%%%%%%%%%%%%%%%%%%%%%%%%%%%%%%%%%%%%%%%%%%%%%
%\begin{figure}[ht]
%\setlength{\epsfxsize}{1.0\hsize}
%\centerline{\epsfbox{fig3.eps}}%\vskip 1.0 cm
\begin{figure}[ht]
{\par
\centering 
\resizebox*{10.9cm}{9.0cm}
{\includegraphics{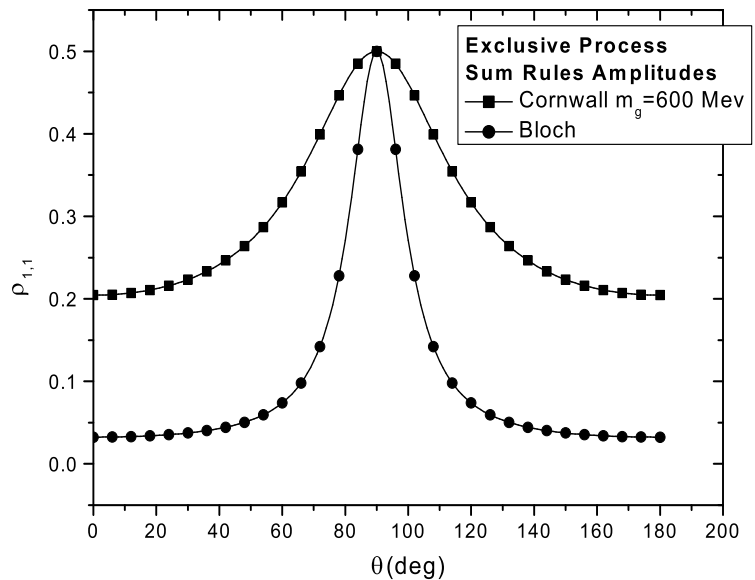}} 
\par}
\caption[dummy0]{Same notation as in Fig.(\ref{exsym}) but calculated with QCD sum
rules distribution amplitudes} \label{exrules}
\end{figure}
%%%%%%%%%%%%%%%%%%%%%%%%%%%%%%%%%%%%%%%%%%%%%%%%%%%%%%%%%%%%%%%

Our results for the exclusive process (Eq.(\ref{exclusive})) are
displayed in Fig.(\ref{exsym}) and Fig.(\ref{exrules}).
Fig.(\ref{exsym}) shows $\rho_{1,1}$ as a function of the
$\rho$-meson production angle $\theta$, computed with the
symmetric wave function whereas Fig.(\ref{exrules}) was calculated
with the QCD sum rule amplitudes. Both figures show curves for
different behaviors of the running coupling constant, the one of
Eq.(\ref{bloch}) obtained by Bloch and the one of Eq.(\ref{acor})
obtained by Cornwall with  $m_g = 600$ MeV and $\L = 300$ MeV.
The calculations for model A are not shown in the figures. The results for
this model are similar to the ones of model B with curves located in the
opposite direction to the curves of model C.

%%%%%%%%%%%%%%%%%%%%%%%%%%%%%%%%%%%%%%%%%%%%%%%%%%%%%%%%%%%%%%%%%%%%
\begin{figure}[ht]
%\setlength{\epsfxsize}{1.0\hsize}
%\centerline{\epsfbox{fig4.eps}}%\vskip 1.0 cm
{\par
\centering 
\resizebox*{10.9cm}{9.0cm}
{\includegraphics{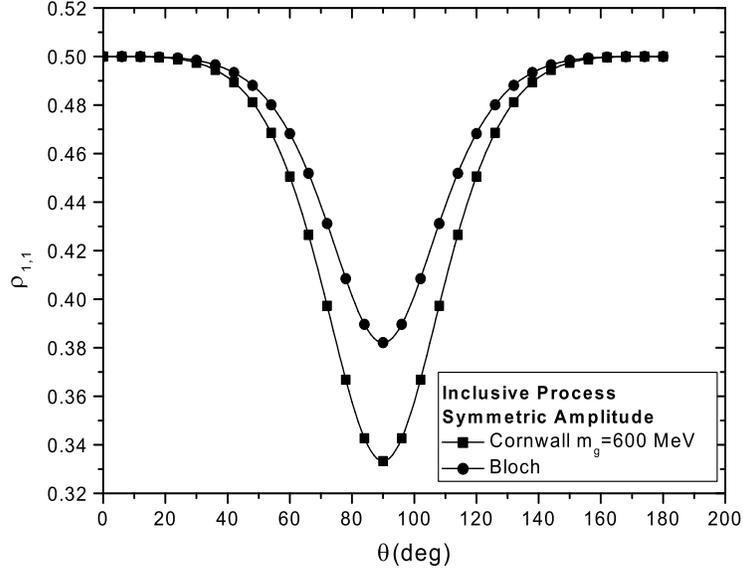}} 
\par}
\caption[dummy0]{Values of the helicity matrix element $\rho_{1,1}$ as a function of
the $\rho$-meson production angle $\theta$ in the case of inclusive $\chi_{c2}$ production.
The curves were obtained using the symmetric distribution amplitudes and the expressions of
Eq.(\ref{bloch}) and Eq.(\ref{acor}) for the coupling constant.} \label{insym}
\end{figure}
%%%%%%%%%%%%%%%%%%%%%%%%%%%%%%%%%%%%%%%%%%%%%%%%%%%%%%%%%%%%%%%

%%%%%%%%%%%%%%%%%%%%%%%%%%%%%%%%%%%%%%%%%%%%%%%%%%%%%%%%%%%%%%%%%%%%
\begin{figure}[ht]
%\setlength{\epsfxsize}{1.0\hsize}
%\centerline{\epsfbox{fig5.eps}}%\vskip 1.0 cm
{\par
\centering 
\resizebox*{10.9cm}{9.0cm}
{\includegraphics{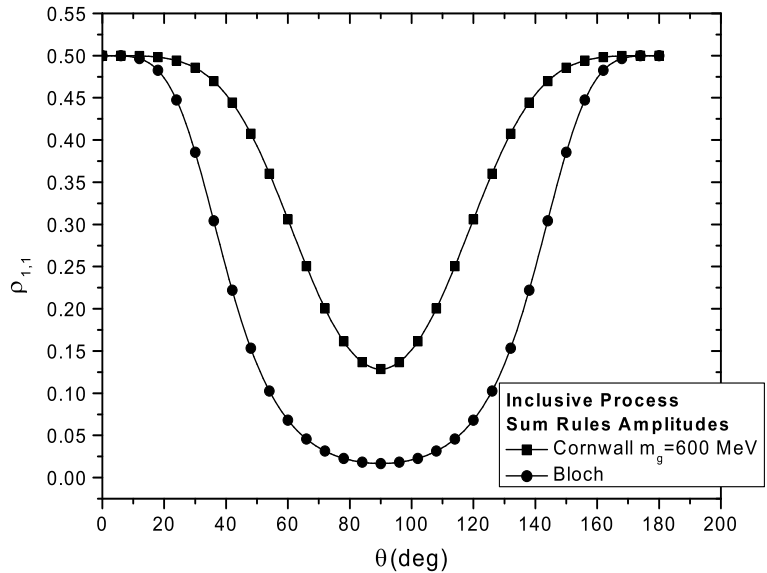}} 
\par}
\caption[dummy0]{Same notation as in Fig.(\ref{insym}) but calculated with QCD sum rules
distribution amplitudes} \label{inrules}
\end{figure}
%%%%%%%%%%%%%%%%%%%%%%%%%%%%%%%%%%%%%%%%%%%%%%%%%%%%%%%%%%%%%%%

As claimed by Anselmino and Murgia \cite{murgia} the measurement of the helicity matrix element
$\rho_{1,1}$ of the vector meson in the $\chi_{c2} \rightarrow \rho\rho$ can indeed discriminate
differences between wave (or distribution) functions, but the effect of different running couplings
is not negligible and one effect can mask the other. Of course, this is not an easy experiment,
but is a feasible one. If the measurement is performed with high precision it can provide
information about the infrared behavior of the coupling constant.

The curves for the inclusive process (Eq.(\ref{inclusive})) are displayed in Fig.(\ref{insym}) and
Fig.(\ref{inrules}). Fig.(\ref{insym}) shows $\rho_{1,1}$ as a function of
the $\rho$-meson production angle $\theta$, computed with the symmetric distribution amplitude whereas
Fig.(\ref{inrules}) was calculated with the QCD sum rule amplitudes. The differences in
the curves are of the same order as in the exclusive case.

Note that in both cases (exclusive or inclusive) the values of $\rho_{1,1}$ is modified
by a factor of $O(2-3)$ when we use the wave functions obtained through QCD sum rules. This is
expected because they have a stronger dependence on $\alpha_s(q^2)$. The difference occurs at different
angles for the exclusive and inclusive $\chi_{c2}$ production. Actually, it is surprising
that in such a delicate experiment, where the effects of the running coupling appear in a
ratio like the one described by Eq.(\ref{ratea}), there is still a clear signal of the
dependence on the infrared value of the coupling constant. In this section we verified
how the infrared behavior of the coupling can be masked by the nonperturbative wave functions.
In the next sections we will look at processes where the effects of wave or distribution
functions are not so pronounced or are better known.

\section{The $\gamma \rightarrow \pi_{0}$ transition form factor}

When making predictions within perturbative QCD, we are always confronted
with problems like the choice of distribution amplitudes as discussed in the
previous section, the choice of renormalization scale $\mu$ and scheme for
the coupling constant, etc... However there are calculations that have been
discussed extensively in the literature, processes where the measurement is
more sensitive only to the asymptotic distribution amplitude of a particular
hadron, and where an optimal renormalization scale has been estimated. This
was the case of the pion form factor discussed in Ref.\cite{we}, and is the
case of the $\gamma \rightarrow \pi_{0}$ transition form factor to be
described here.

The photon-to-pion transition form factor $F_{\gamma\pi}(Q^2)$ is measured in
single-tagged two-photon $e^+e^- \rightarrow e^+e^- \pi^0$ reactions. The
amplitude for this process has the factorized form

\begin{equation} \label{amppi}
F_{\gamma\pi}(Q^2) = \frac{4}{\sqrt{3}} \int^1_0 \, dx \, \phi_{\pi}(x,Q^2) T^H_{\gamma\pi}(x,Q^2),
\end{equation}

\noindent
where the hard scattering amplitude $ T^H_{\gamma\pi}(x,Q^2)$ is given by

\begin{equation} \label{hardgpi}
T^H_{\gamma\pi}(Q^2) = \frac{1}{(1-x)Q^2} [ 1 + {\cal O}(\alpha_s)].
\end{equation}

\noindent
Using an asymptotic form for the pion distribution amplitude $\phi_{\pi}=\sqrt{3} f_\pi x(1-x)$
we obtain \cite{brodsky}

\begin{equation}\label{trans}
Q^2F_{\gamma\pi}(Q^2)=
2f_{\pi}\left(1-\frac{5}{3}\frac{\alpha_{V}(Q^\ast)}{\pi}\right)
\end{equation}

\noindent
where $Q^\ast = \exp^{-3/2} Q$ is the estimated Brodsky-Le\-pa\-ge-\-Mac\-ken\-zie scale
for the pion form factor in the scheme discussed in Ref.\cite{brodsky}.

\begin{figure}[ht]
%\setlength{\epsfxsize}{1.0\hsize}
%\centerline{\epsfbox{fig6.eps}}%\vskip 1.0 cm
{\par
\centering 
\resizebox*{12.14cm}{10.0cm}
{\includegraphics{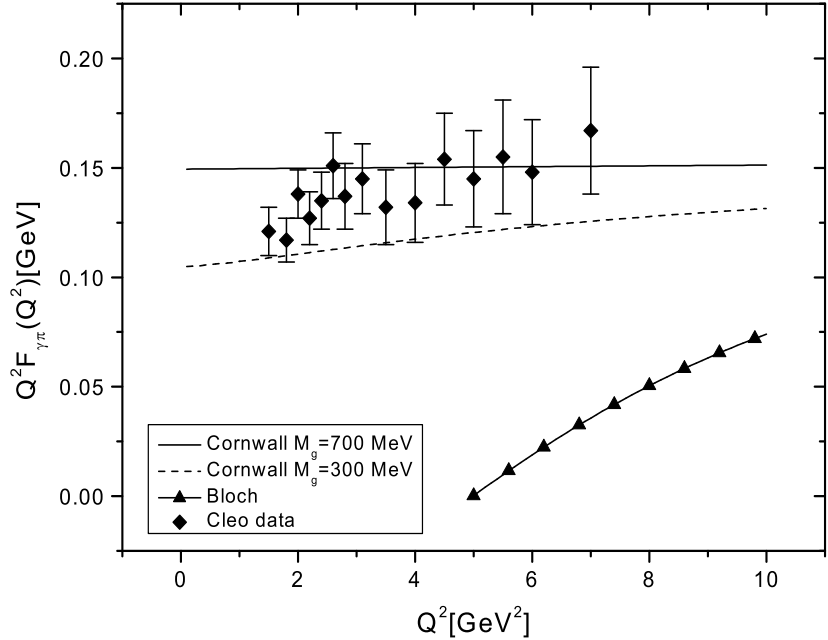}} 
\par}
\caption[dummy0]{The $\gamma \rightarrow \pi^{0}$ transition
form factor calculated with different expressions for the infrared
behavior of the running coupling constant.} \label{transition}
\end{figure}

In Fig.(\ref{transition}) we compare the photon to pion transition form factor
with CLEO data \cite{cleo}. The curves were computed with different expressions for the
infrared behavior of the running coupling constant. We assumed
$f_{\pi} \simeq 93 \;\mbox{MeV}$ and  $\L = 300$ MeV. Using the running coupling constant
given by the expression of Eq.(\ref{bloch}) we obtain a fit for the photon-pion transition
form factor very far from the experimental data. The result obtained when we use Eq.(\ref{runalk})
is not shown and gives an even worse fit. The infrared value of the coupling
constant is so large in the case of the coupling constants given by Eqs.(\ref{bloch}) and (\ref{runalk}),
that we are not sure that the perturbative result can be trusted even at such large momentum scale.

%%%%%%%%%%%%%%%%%%%%%%%%%%%%%%%%%%%%%%%%%%%%%%%%%%%%%%%%%%%%%%%%%%%%
%%%%%%%%%%%%%%%%%%%%%%%%%%%%%%%%%%%%%%%%%%%%%%%%%%%%%%%%%%%%%%%

The infrared coupling constants related to the class of SDE solutions consistent with
an infrared finite propagator that vanishes at origin of momenta, are much stronger
than most of the phenomenological estimates of the frozen $\alpha_s(0)$ value
that we quoted in Ref.\cite{we} ($\alpha_s (0) \approx 0.7 \pm 0.3$),
and are at the origin of the strange lower curve of Fig.(\ref{transition}). Notice
that due to the large scale of momenta involved in this process it is not expected
that higher order corrections are still important. The data is only compatible with
Eq.(\ref{acor}), which has a smoother increase towards the infrared region. Perhaps
this behavior is actually indicating that the transition to the infrared should be
a soft one.  Again the result for model A is not shown because it is even worse than the
one of model B.

Note that in Fig.(\ref{transition}) the curves obtained with the
Cornwall's coupling constant do not show large variation in the full range of
uncertainty of the dynamical gluon mass (see Eq.(\ref{mg})). It is interesting that
its behavior is quite stable in this case as well as for the pion form factor
studied in Ref.\cite{we}. If we had large variations of the infrared coupling
constant with the gluon mass scale we could hardly propose any reliable
phenomenological test for its freezing value. We stress that the results are
obtained for a perturbative scale of momenta and it seems that we have to choose
between two possibilities: This process cannot be predicted by perturbative
QCD up to a scale of several GeV, or some of the ESD solutions are predicting a
too large value of the coupling constant in the infrared and the approximations
made to determine these solutions are too crude.

\section{Elastic differential cross section for pp scattering}

We have verified in Section IV one case where the behavior of the
coupling constant in the infrared is entangled with the one of the distribution
amplitudes. In Section V it was shown a classic QCD calculation where
the asymptotic behavior of the wave function is more important than the infrared one,
leading to a cleaner test for the nonper\-tur\-ba\-tive behavior of the coupling constant. Here we
will discuss a test for the infrared QCD behavior that makes use of a phenomenological
model for diffractive interactions \cite{halzen}. In this case we will also make use
of a hadronic wave function, but the main source of uncertainty is in the fact that we assume a model
for elastic scattering containing a series of approximations, which are beyond the scope
of a pure QCD calculation.

We will compute the elastic differential cross section for pp scattering within
the Landshoff-Nachtmann model (LN) for the Pomeron exchange \cite{land}. As discussed in
Refs.\cite{halzen,che}, this is a model where the Pomeron is represented by two-gluons
exchange, and it is particularly dependent on the infrared properties of the gluon, not
only on the coupling constant but also on the behavior of the gluon propagator. In this
model one of the gluons carry most of the momentum exchanged in the interaction while the
other seems to appear just to form the colorless Pomeron. Therefore
besides the coupling constant we will also need the expressions for the gluon propagators
obtained through the solutions of the SDE, although this dependence with the nonperturbative
behavior of the gluon propagator is important only when the gluon is exchanged at
low momentum. The gluon propagator in Landau gauge is written
as

\begin{equation} D_{\mu\nu}(q^2)= \({\delta}_{\mu\nu}
-\frac{q_{\mu}q_{\nu}}{q^2}\)D(q^2),
\label{landau}
\end{equation}

\noindent
where the expression for $D(q^2)$ obtained by Cornwall is given by

\begin{equation}\label{propcorn}
 D^{-1}(q^2) = \left[q^2 + M_g ^2(q^2)\right]bg^2\ln\left[\frac{q^2+ 4M^2_{g}}{\Lambda ^2}  \right].
\end{equation}

\noindent
We will also make use of the running coupling constant obtained by Fischer and
Alkofer (model A) and their respective propagator $ D(q^2) = Z(q^2)/q^2$,
where $Z(q^2)$, in Landau gauge, is fitted by

\begin{equation}\label{propgalk}
  Z(x)= \left(\frac{\alpha_{sA}(x)}{\alpha_A(\mu)}\right)^{1+ 2\delta}R^2(x),
\end{equation}

\noindent
and

\begin{equation}\label{rpropg}
R(x)= \frac{cx^{\kappa}+ dx^{2\kappa}}{1 + cx^{\kappa}+
dx^{2\kappa} }
\end{equation}

\noindent
where the constants appearing in Eq.(\ref{propgalk}) and Eq.(\ref{rpropg}) are given by

\hskip 0.5in $\alpha_A (\mu^2) = 0.9676$,

\hskip 0.5in $\kappa= 0.5953$,

\hskip 0.5in $\delta=-9/44$,

\hskip 0.5in $c= 1.8934\;\mbox{GeV}^{-2\kappa}$,

\hskip 0.5in $d= 4.6944\;\mbox{GeV}^{-4\kappa}$.

In the LN model the elastic differential cross can be obtained from

\begin{equation}\label{dsigma}
  \frac{d\sigma}{dt}= \frac{|A(s,t)|^2}{16\pi s^2}
\end{equation}

\noindent
where the amplitude for elastic proton-proton scattering via
two-gluon exchange can be written as

\begin{equation}
A(s,t)= is8\alpha_{s}^2\left[T_{1} - T_{2}\right]
 \label{ampli}
\end{equation}

\noindent
with

\begin{equation}
T_{1}= \int\,d^{2}k
D\(\frac{q}{2}+k\)D\(\frac{q}{2}-k\)|G_{p}(q,0)|^{2} \label{t1}
\end{equation}

\begin{eqnarray}
T_{2}= \int\,&&d^{2}k D\(\frac{q}{2}+k\)D\(\frac{q}{2}-k\)G_{p}
\(q,k-\frac{q}{2}\)  
%\nonumber \\
%&&\times
\left[2G_{p}(q,0)-G_{p}\(q,k-\frac{q}{2}\)\right ]
\label{t2}
\end{eqnarray}

\noindent
where $G_p (q,k)$ is a convolution of proton wave functions

\begin{equation}
G_p(q,k)= \int \, d^2pd\kappa \psi^\ast(\kappa,p)\psi(\kappa,p-k-\kappa q).
\label{gdp}
\end{equation}

\noindent
$G_p (q,0)$ is given by the Dirac form factor of the proton

\begin{equation}
F_{1}(t) = G_p(q,0)= \frac{4m^2 - 2.79t}{4m^2
-t}\frac{1}{(1-t/0.71)^2} \label{dirac} .
\end{equation}

To estimate $G_p (q,k-q/2)$ we assume a proton wave function peaked at
$\kappa = 1/3$ and obtain

\begin{equation}
 G_p\(q,k-\frac{q}{2}\)=F_{1}\( q^2 + 9\left|k^2-\frac{q^2}{4}\right| \)
\label{conv}
\end{equation}

Eq.(\ref{dsigma}) to Eq.(\ref{conv}) can be computed using the couplings
and propagators discussed above. We compare the differential elastic cross-section for
proton-proton scattering with the experimental data of Breakstone {\it et al.} at
$\sqrt{s}= 53 \quad {\textnormal GeV}$ \cite{breakstone}. The results have
to be adjusted by a normalization factor $s^{0.168}$ which accounts for the
energy dependent part of ${d\sigma}/{dt}$ \cite{land2}. For large $|t|$ values,
double-Pomeron exchange and three-gluon exchange are known to be important;
we therefore do not expect to describe the data near and above
$-t \simeq 0.5 \quad {\textnormal GeV}^2$ \cite{land2,don}

\begin{figure}[ht]
%\setlength{\epsfxsize}{0.9\hsize}
%\centerline{\epsfbox{fig7.eps}}%\vskip 1.0 cm
{\par
\centering 
\resizebox*{10.9cm}{9.0cm}
{\includegraphics{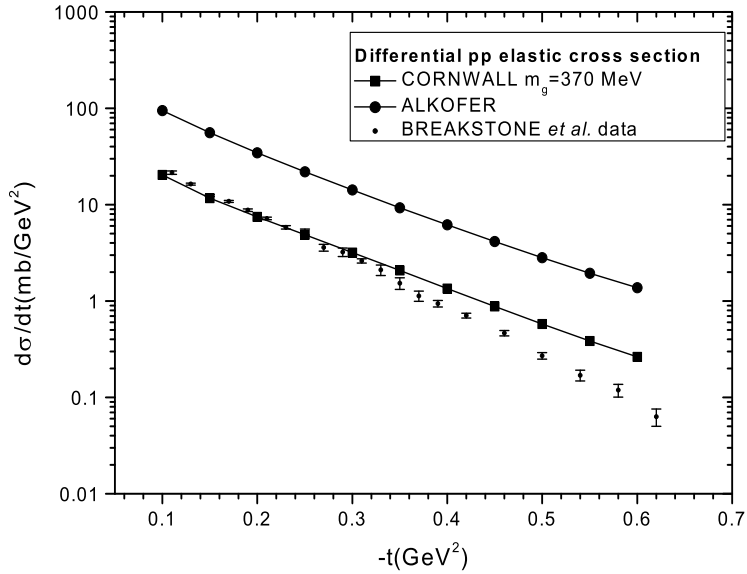}} 
\par}
\caption[dummy0]{ Differential $pp$ elastic cross section at
$\sqrt(s)=53\, \mbox{GeV}$ computed within the
Landshoff--Nachtmann model for the Pomeron,
using different infrared couplings and gluon propagators
obtained from SDE solutions. }
\label{cross-sec}
\end{figure}

The data of elastic proton-proton scattering is well fitted assuming a dynamical
gluon mass of $m_g = 370 \quad {\textnormal MeV}$ for $\L= 300 \quad {\textnormal MeV}$.
The calculation has a small variation with the value of $m_g$ and is more sensitive
to the ratio $m_g /\L$. Using the coupling and propagator of Alkofer {\it et al.} (model A) we obtain
a curve that is about one order of magnitude away from the experimental points.
Unfortunately we do not have the propagator solution for model B and for this reason we
do not show the results in this case, although we can guess that it would produce a curve
in Fig.(\ref{cross-sec}) between the curves of model A and C.

%%%%%%%%%%%%%%%%%%%%%%%%%%%%%%%%%%%%%%%%%%%%%%%%%%%%%%%%%%%%%%%%%%%%
%%%%%%%%%%%%%%%%%%%%%%%%%%%%%%%%%%%%%%%%%%%%%%%%%%%%%%%%%%%%%%%%

There is a striking difference between the two classes of SDE solutions
that we discussed in the Introduction. In the case of Cornwall's solution it is
the product $g^2 D(q^2)$ that behaves roughly as $1/{m_g}^2$ as $q^2 \rightarrow 0$ and
this product has no $g^2$ dependence \cite{cornwall}. This does not happen for the
class of gluon propagators that seems to vanish at origin. As in this case the coupling is not
canceled in the ${d\sigma}/{dt}$ calculation and as it is a factor of $O(3)$ larger,
it is not difficult to understand the one order of magnitude difference in the
result of the cross sections. This specific calculation is model dependent, however
it is quite sensitive on the infrared behavior of the theory.

\section{Exclusive $\rho$ Production in Deep Inelastic Scattering}

Donnachie and Landshoff \cite{dl1} successfully described the process
$ \gamma ^{*}(q)\: p(P)\: \rightarrow \: \rho (r)\: p(P^{\prime} ) $ with a soft Pomeron
exchange, and later they \cite{dl2} considered Pomeron exchange as being
the exchange of two gluons as in the previous section.
Pursuing that idea Cudell \cite{cudell} proposed
the following expression for the polarized differential cross section for exclusive
$\rho$ production:
\begin{equation}
\label{eq:dsdt_pol}
\frac{d\sigma _{j}}{dt}=\left( \frac{\alpha _{elm}}{4w^{4}}\left| A_{j}\right| ^{2}\Phi ^{2}\right) \: Z^{2}\:
\left[ 3\: F_{1}(t)\right] ^{2}\: ,\quad j=T,\: L ,
\end{equation}
where $ \alpha _{elm}\simeq 1/137 $ is the electromagnetic coupling constant
and $ w^{2}=(q+P)^{2} $. The factor $\Phi$ is given by
\begin{equation}
\label{eq:phi}
\Phi =\sqrt{\frac{f_{\rho }\: m_{\rho }}{24}}  ,
\end{equation}
where $ f_{\rho }\simeq 30 $MeV is the $ \rho  $ form factor and $ m_{\rho }\simeq 770 $ MeV
is the $ \rho  $ mass.
\begin{equation}
\label{eq:Z}
Z=\left( \frac{w^{2}}{w_{0}^{2}}\right) ^{0.08+\alpha^{\prime} t}\: ,
\end{equation}
 with $ w^{2}_{0}=1/\alpha^{\prime} \simeq 4 $GeV$ ^{2} $. The Dirac form
factor of the proton is
\begin{equation}
\label{eq:F1}
F_{1}(t)=\frac{4m^{2}_{p}-2.79t}{4m^{2}_{p}-t}\frac{1}{(1-t/0.71)^{2}},
\end{equation}
 where $ m_{p} $ is the proton mass.

The amplitude is
\begin{eqnarray}
\label{eq:Aj}
&&A_{j}=i\frac{8\sqrt{2}}{3\pi }\: m_{\rho }\: P_{j}\:                      
%%\\
\int^{0}_{-\infty}
\frac{dk^{2}\; (t-4k^{2})}{(\mu^2_Q - t)(\mu^2_Q - 4k^{2})}
\left[4\pi \alpha _{n} D\left(k^{2}+\frac{t}{4}\right)\right]^{2}, \nonumber\\
&&\quad\quad\mu^2_Q = m^{2}_{\rho}+Q^{2}\;,\quad j = T,\: L \nonumber
\end{eqnarray}
where $P_{T}=w^{2}/2 $; $ P_{L}=P_{T}\cdot (m^{2}_{\rho }+Q^{2}+t)/(2m_{\rho }Q) $.
In the last term of the integrand we have the frozen coupling constant $ \alpha _{n}=\alpha _{s}(0) $
and the gluon propagator $ D(p^{2}) $.
\begin{figure}[ht]
{\par\centering \resizebox*{10.9cm}{9.0cm}{\includegraphics{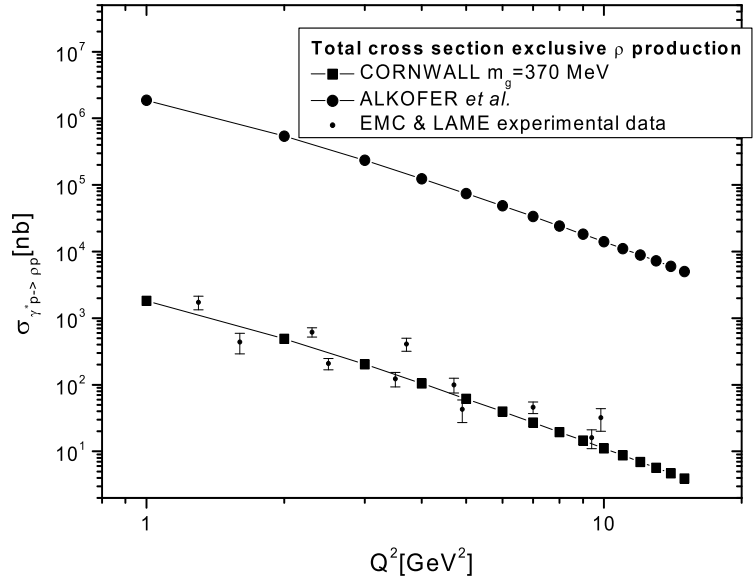}} \par}
\caption{Total cross section for exclusive $\rho$ production in deep inelastic scattering.}
\label{fig:rhosigmas}
\end{figure}
The total cross section can be obtained through the integral
\begin{equation}
\label{eq:sigtotal}
\sigma _{total}(Q^{2})=\int dt\: \left( \frac{d\sigma _{T}}{dt}+\varepsilon \frac{d\sigma _{L}}{dt}\right)
\: ,\quad \varepsilon \approx 0.85.
\end{equation}

This cross section is dependent not only on the nonperturbative coupling
constant but also on the gluon propagator as the calculation of the previous section.
We compared the solutions of Alkofer \textit{et al.} (model A) and Cornwall (model C). The reason
for not using model B is the same as the one discussed in the previous section.
With the running coupling constants (eqs.~(\ref{runalk}) and (\ref{acor})) and
their respective propagators (eqs.~(\ref{propgalk}) and (\ref{propcorn}))
we obtain the results shown in Fig.(\ref{fig:rhosigmas}),
where they are compared with the experimental data
from EMC\cite{emc} and LAME\cite{lame}.
As explained in Ref.\cite{cudell} the model for exclusive $\rho$ production is consistent with experiment
only in the range of momenta shown in the figure, at larger values of $Q^{2}$ there are other contributions
to the cross section.

\section{Conclusions}

In this work we proposed phenomenological tests for the infrared behavior of
running coupling constant that have been obtained through the nonper\-tur\-ba\-tive
solutions of Schwinger-Dyson equations of the QCD gluonic sector. We are only
considering SDE solutions that are infrared finite. Although infrared divergent
solutions have not been fully discarded, there are several indications that these
QCD Green functions are well behaved in the infrared. Unfortunately
due to the complexity of the SDE, approximations are still necessary to obtain
these solutions and they lead to different expressions.
Phenomenological tests are important because they are able to
indicate which are the correct approximations that can be performed to solve SDE,
selecting the solution that is compatible with the experimental data.

We studied the effects of a frozen coupling constant in the $\tau$-lepton decay
rate into nonstrange hadrons. This is a typically perturbative test and, in particular,
it shows the importance of the method used to fix the scale of the running coupling constant,
which is different for the couplings shown in Eqs.(\ref{runalk}) and (\ref{bloch}) (model
A and B). The measurement of the $\rho$
vector meson helicity density matrix that are produced in the $\chi_{c2}\rightarrow
\rho\rho$ decay is one interesting experiment to be performed.
The measurement of the diagonal matrix element provide information
on the infrared behavior of the coupling constant as well as on the distribution
amplitude. In this case both behaviors are entangled and a good knowledge on the
distribution amplitude is needed in order to select a preferred behavior for
the coupling constant. The experiment is not an easy one, but can be performed
and will give information complementary to the others that we have discussed.

The effects of the infrared behavior of the coupling constant in the photon-to-pion
transition form factor $F_{\gamma\pi}(Q^2)$ is much more interesting. The calculation
probes the asymptotic behavior of the pion distribution amplitude and provide a
cleaner test for the freezing of the coupling constant. The differences between
the existent solutions obtained from SDE for $\alpha_s (q^2)$ are shown clearly in
the comparison with the experimental data, which seems to select the coupling
constant compatible with dynamical generation of a gluon mass.

The analysis of the differential cross section for proton-proton scattering and the
cross section for exclusive $\rho$ production in deep inelastic
scattering are also unambiguously selecting one specific behavior for the QCD Green functions
in the infrared. In this case we made use of a model of diffractive scattering
proposed by Landshoff and Nachtmann and it could be argued that the model has not
been obtained from straightforward QCD calculations. However the model has a great
success in the explanation of diffractive physics at low energy, and again the
comparison with experimental data is consistent with the result obtained from
the other phenomenological tests. Diffraction in this model is explained through
the exchange of two gluons, where one of the gluons carries most of the momentum
and the other is soft. Therefore the model actually probes the infrared region.

It is worth mentioning the influence of gauge invariance and presence of fermions
in the solutions for the coupling constants that we have discussed. The Cornwall's
solution was obtained in a gauge invariant procedure, the others were obtained in
Landau gauge. We do not expect that gauge invariance introduce a large effect in the
result. In general solutions of SDE are relatively stable in respect to changes
in the gauge choice, and we should expect that if the solutions minimize the vacuum
energy the gauge dependence should disappear. The effect of the number of flavors
in Cornwall's solution\cite{cornwall} is not so strong, and it appears in the
coefficient $\beta_0$ of the coupling constant and in the gluon
mass equation increasing the value of the frozen coupling. If a
nonzero number of flavors produces any observable effect, this
one should act in the same sense for all solutions. Therefore,
we do not expect large changes in our results with the inclusion
of fermion loops in the SDE solutions.

We conclude pointing out that Schwinger-Dyson equations provide a powerful tool to
investigate the QCD infrared behavior. These are complicated equations whose solutions
are determined only after some specific approximations leading to different expressions.
These expressions can be phenomenologically tested, indicating which are the most
reliable approximations. In the tests presented here only one solution, consistent
with a dynamically generated mass for the gluon, was shown to be compatible with the
experimental data.

\section*{Acknowledgments}

This research was supported by the Conselho Nacional de
Desenvolvimento Cient\'{\i}fico e Tecnol\'ogico (CNPq) (AAN), by
Funda\c{c}\~ao de Amparo \`a Pesquisa do Estado de S\~ao Paulo
(FAPESP) (ACA,AAN) and by Coordenadoria de Aperfei\c coamento do
Pessoal de Ensino Superior (CAPES) (AM).

\end{document}